\documentclass[letterpaper]{article}
\usepackage{aaai}
\usepackage{times}
\usepackage{helvet}
\usepackage{courier}

\usepackage{graphicx}
\usepackage{soul}

\begin{document}
%
\title{Scopes of Alignment}
\author{Kush R. Varshney, Zahra Ashktorab, Djallel Bouneffouf, Matthew Riemer, and Justin D. Weisz \\ IBM Research \\ 1101 Kitchawan Road\\ Yorktown Heights, NY 10598\\}

\maketitle
\begin{abstract}
\begin{quote}
Much of the research focus on AI alignment seeks to align large language models and other foundation models to the context-less and generic values of helpfulness, harmlessness, and honesty. Frontier model providers also strive to align their models with these values. In this paper, we motivate why we need to move beyond such a limited conception and propose three dimensions for doing so. The first scope of alignment is \emph{competence}: knowledge, skills, or behaviors the model must possess to be useful for its intended purpose. The second scope of alignment is \emph{transience}: either semantic or episodic depending on the context of use. The third scope of alignment is \emph{audience}: either mass, public, small-group, or dyadic. At the end of the paper, we use the proposed framework to position some technologies and workflows that go beyond prevailing notions of alignment.
\end{quote}
\end{abstract}

\section{Introduction}

A new paradigm of artificial intelligence (AI) is upon us. Large language models (LLMs) or foundation models, to refer to the same concept irrespective of data modality more generally, are finding uses across application domains \cite{bommasani2021opportunities}. Such use is often to generate new outputs---such as text, images, code, or molecules---based on the training data and a prompt. Although there is excitement for LLMs, and practical applications are seeing a return on investment, we must be aware of their risks and harms \cite{weidinger2022taxonomy,shelby2023sociotechnical}. Traditional risks of AI, like lack of fairness, robustness, explainability, transparency, and uncertainty quantification are present in LLMs, but LLMs amplify the existing harms and introduce new harms. Some new harms include hallucination or lack of factuality, hateful speech, prompt injection, information leakage, copyright infringement, bullying and gaslighting. Thus, a major current endeavor in responsible AI beyond making LLMs as helpful as possible is to make them as harmless and honest as possible \cite{sun2024trustllm}. LLMs are being described in terms of their behaviors using the language of psychology. Although we must be careful with respect to anthropomorphism \cite{ShneidermanM2023}, behavior is emerging as the lingua franca for specifying desired input--output relationships of LLMs and for evaluating their performance. 

\begin{figure*}
  \centering
  \includegraphics[width=1.4\columnwidth]{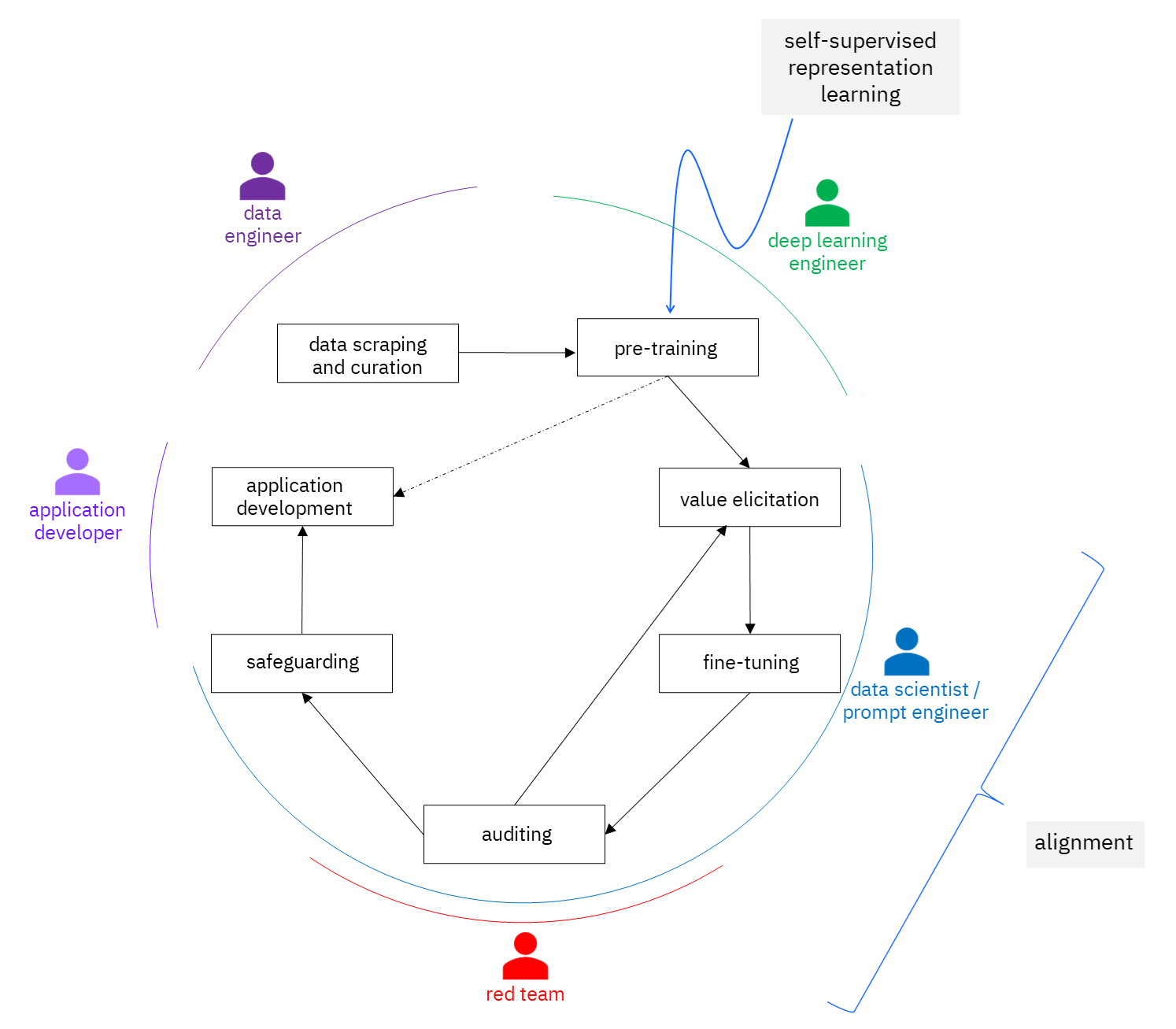}
  \label{fig:lifecycle}
  \caption{Typical application development lifecycle for LLMs considered by frontier model providers.}
\end{figure*}

\emph{Values} are fundamental beliefs that guide behaviors. They indicate the importance of various things and actions to a person or group of people, and determine the best ways to behave. As shown in Fig.~\ref{fig:lifecycle}, there are many ways throughout the AI application development lifecycle to get an LLM to behave in a desired fashion. Data scraping, curation and pre-training an entire model from scratch offer an approach that is often prohibitively costly in terms of both data and computation. Safeguarding, through moderations or constrained decoding based on classifiers or detectors for various harm dimensions \cite{achintalwar2024detectors}, is lightweight in data and computing resources but may not entirely capture all sorts of desired behaviors. 

The steps in the middle of the lifecycle between LLM pre-training and safeguarding are known these days as \emph{alignment} \cite{shen2023large}. Alignment steps are meant to embed behaviors and values in the AI system and may be done over more than one round. (In the figure, the dotted line implies that alignment is an optional step.) However, alignment is an `empty signifier': something without an underlying meaning \cite{kirk2023empty}. Different parties have appropriated the term to refer to various goals and actions. Lipton quips \cite{lipton2024}: ``Alignment is now defined so broadly that all of AI, all of ML, and the entire history of technology is---and always has been---`alignment research'.'' Gilbert quips \cite{gilbert2024hcc}: ``Alignment---the sensible kind anyway---is just human-centered computing.'' Frontier model providers typically refer to alignment as endowing LLMs with instruction-following behavior and helpful/harmless/honest values \cite{Bai2022,Ouyang2022}.

There are several ways of carrying out the goal of alignment, each having a different cost profile. Full fine-tuning changes all of the parameters of the LLM and can be done through supervised learning and/or reinforcement learning. Parameter efficient fine-tuning changes a small number of parameters, often through low-complexity adapters. Prompt engineering does not change the LLM at all, but relies on prompts to achieve the desired behavior. Additionally, system prompts --- special prompts appended before all inputs --- are commonly developed and sometimes made public by frontier model providers \cite{anthropic2024system}.

\section{Limitations of Alignment}

There are limitations to the LLM lifecycle and how LLM alignment has been conceived and carried out to date. The alignment methods and approaches of frontier model providers lead to a singular set of values that are intended across all contexts, all deployments, and all time. Many recent empirical studies show the set of knowledge and values within LLMs is sociopolitically biased toward dominant cultural sources \cite{Durmus2023,Feng2023}. The root cause is alignment, as much as, or more than pre-training \cite{qi2023fine}. 

Moral psychology is the key sub-area within psychology that deals with values. Morality is the differentiation of behaviors between those that are right (proper) and those that are wrong (improper); it can be a body of principles derived from a particular religion or culture. It can be highly personal as well. Morality captured by multi-lingual language models does not reflect cultural differences, but is dominated by high-resource languages and cultures \cite{haemmerl-etal-2023-speaking,scherrer2024evaluating}. Such analysis has been organized around moral foundations theory, which sets up six dichotomies: care/harm, fairness/cheating, loyalty/betrayal, authority/subversion, sanctity/degradation, and liberty/oppression \cite{graham2013moral}. 

Clear patterns show up empirically through the lens of moral foundations theory, but we may ask if there is something more basic going on. The prevailing approach to alignment seems to espouse the silver rule --- do not do to others what you would not like them to do to you --- and the golden rule --- do to others what you would like them to do to you. The focus is completely on the person doing the action (agent) and their perception, not on the person to whom the action is done (patient). In contrast, the platinum rule that foregrounds the patient and their perception --- do to others what \emph{they} would like done to them --- is currently not adequately incorporated in alignment. 
Yet, all three are important.

Dyadic morality theory, an alternative to moral foundations theory, specifically considers moral issues as situations where an intentional agent is perceived to cause damage to a vulnerable patient \cite{schein2018theory}. And, it is the differences in the perception of who or what is an intentional agent, who or what is a vulnerable patient, and what constitutes a causal relationship, that lead to differences in morality. Are weather phenomena intentional agents? How about AI systems? Are unborn children vulnerable patients? Can the saying of a mantra be the cause of damage? Moreover, these perceptions may be different and change based on the type of damage, the context, or other factors. Thus, dyadic morality theory is less universal and more contextually-scoped than the moral foundations theory that undergirds prevailing views of alignment.

It is under this light of different perceptions of intentional agents causing damage to vulnerable patients that we state the need for different scopes of alignment, i.e.\ different strokes for different folks. We emphasize that this call for different scopes of alignment is a precursor to dealing with \emph{conflicting} values, skills and knowledge that is more often considered in the nascent field of pluralistic alignment \cite{Dognin2025}. We believe it should be dealt with first before going to the harder problem of simultaneously conflicting values. In fact, by scoping the alignment properly, we may even avoid the conflicts that arise in larger scopes.

\section{Different Scopes of Alignment}

Alignment should not be just one activity, technical approach, or workflow. It should be more specific and scoped for the different needs of different groups over time. Toward this end, we propose the following three scopes of alignment: (1) competence, (2) transience, and (3) audience.

By the \emph{competence} of alignment, we mean which kind of capability the LLM is being given further instruction or fine-tuning in: knowledge, skills, or behaviors. Knowledge is a topically-organized set of facts and information that supports work-related performance. Skills --- what the model can do --- are about the model's proficiency and ability to perform a job-related activity that contributes to effective performance. Some examples of skills displayed by LLMs include summarizing emails and converting texts into hip-hop raps. Behaviors are the values, attitudes, and temperament evidenced through the actions of LLMs. Behaviors can include verbosity, politeness, succinctness, objectivity, formality, inclusiveness, respecting social norms, non-egregiousness, faithfulness, and non-profanity. In human terms, the United States Army, as part of its Army Talent Attribute Framework, counts many different knowledge, skills, and behaviors (KSBs) as essential attributes that ``allow individuals to achieve goals of greater complexity and scale than they can achieve on their own'' \cite{homer2023soldier}. Presumably they would also be essential attributes for AI systems to team with people to achieve their goals. 

By the \emph{transience} of alignment, we mean the boundedness to time or context. Using terminology stemming from neuroscience, we posit two categories of alignment transiences: \emph{episodic} and \emph{semantic}. Episodic alignment is for particular knowledge, skills or behaviors bound to a time, place, or other context. Semantic alignment is for general KSBs about the world that applies to all times and contexts. Semantic knowledge includes concepts like an understanding of the law of gravitation, whereas episodic knowledge may include the state of things in a dynamic environment, such as who is in the building right now. Semantic behaviors may include general duties that are valued across contexts, such as not stealing. In contrast, episodic behaviors may include particular behaviors from an organization's code of conduct, such as obtaining manager pre-approval before booking air tickets, or ones that are contextual based on relational ethics, such as an attorney not divulging incriminating conversations with their client. The behavior transience dichotomy of LLMs is described using the terminology s\={a}dh\={a}ra\d{n}a-dharma and vi\'{s}e\d{s}a-dharma by \cite{Varshney2024}. Skills may be similarly dichotomized. 

\begin{figure*}
  \centering
  \includegraphics[width=1.8\columnwidth]{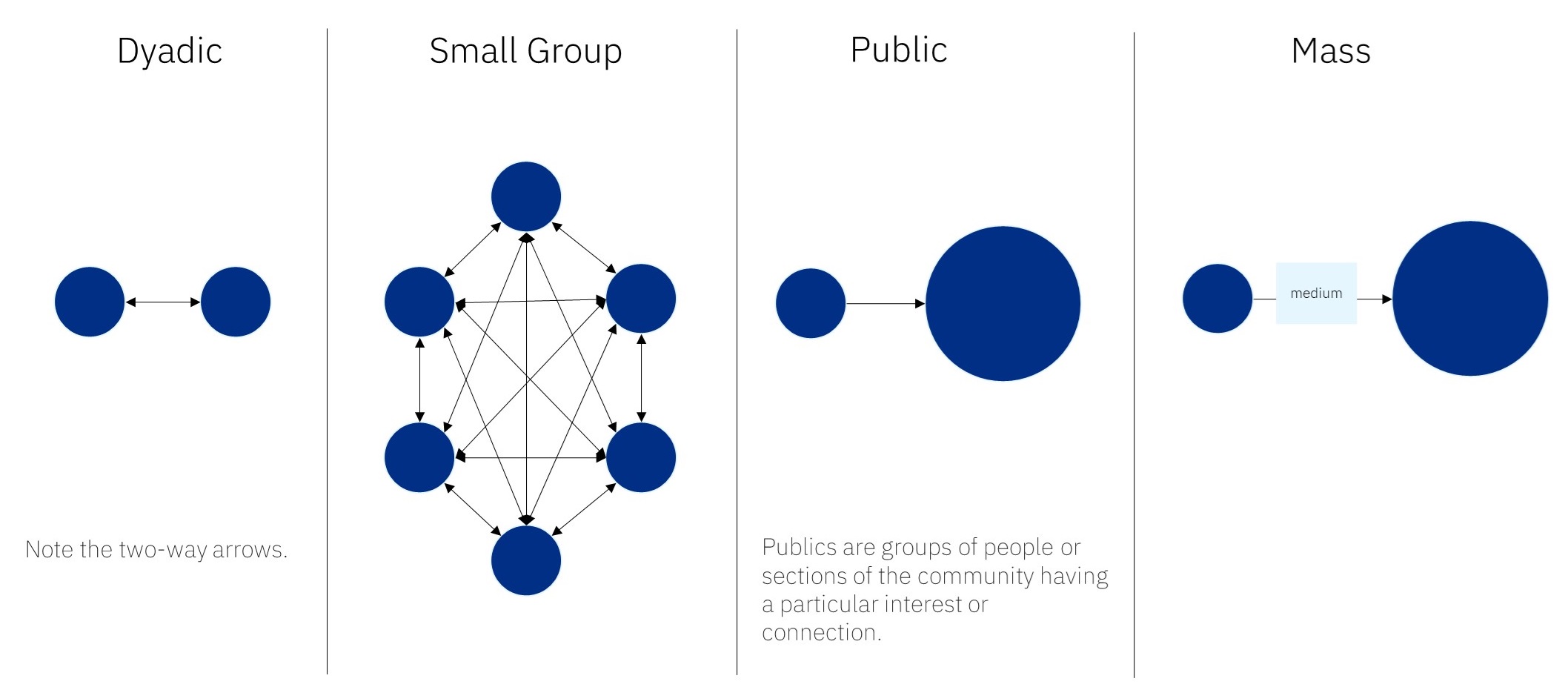}
  \caption{Different audiences of alignment.}
  \label{fig:audience}
\end{figure*}

By the \emph{audience} of alignment, we mean what kind of group the alignment, and ultimately the deployed model, is for. Taking a cue from communication theory, groups can range from a single person interacting with an AI system like an AI personal assistant or therapist (dyadic), to a small group, to a community (public), to everyone (mass). As illustrated in Fig.~\ref{fig:audience}, dyadic alignment between one person and one AI system is a two-way street: people and the LLM can both adapt to achieve better alignment. In recent human-computer interaction literature, this idea has been referred to as mutual theory of mind \cite{wang2024theory,wang2022mutual}.
This can also be the case in a small-group environment with a handful of people and one or more LLM-based agents. However, public alignment and mass alignment are generally focused on the model being aligned to the needs of a community or all of humanity. The medium that distinguishes public alignment from mass alignment is analogous to the difference between simply speaking to a community in-person and requiring media such as newspapers, radio, television, or social networks to broadcast a message. In the case of mass LLM alignment, the communication is in the opposite direction and the medium needs to somehow bring in the valued knowledge, skills, and behavior from all individuals, groups, communities, cultures, and publics. It must be a representation of the full diversity of human culture without missing anything. We do not know of any such effective medium.

\section{Motivating Example}
A compelling example for demonstrating the need for different scopes of alignment is a (hypothetical) AI system designed to support mental health counseling in the United States and China, where cultural norms and regulatory environments differ significantly.

Competence: In the United States, where openness and individualized therapy are emphasized, AI alignment would focus on competencies such as empathic communication, offering personalized strategies, and handling a wide array of mental health issues with flexibility. In China, however, AI competence may need to align with collectivist values, prioritizing solutions that emphasize family and community-oriented approaches, as well as deference to social harmony and privacy regulations.

Transience: In the U.S., episodic alignment might be prioritized, allowing the AI to adapt dynamically to diverse individual needs and new mental health practices as they emerge. In China, semantic alignment may be favored, as the AI system aligns more consistently with established practices and cultural norms around mental health, which are more stable and slower to change in response to new trends.

Audience: The audience of alignment would also vary. In the U.S., the AI system might be tailored for individuals in private settings, including direct interactions with therapists or mental health apps. In China, the audience of alignment could target both individuals and local community health groups, with the AI’s output possibly needing to be accessible to family members or other community stakeholders in keeping with collectivist support structures.

This example illustrates why a flexible, multi-scope approach is essential for AI alignment, as country-specific social and cultural norms shape how AI should interact, evolve, and serve its intended audience.

\section{Implications}

What is the use of defining these three scopes of alignment? The first use is simply to be able to more precisely discern that alignment is not a singular process; it contains many possibilities. Most of the literature focuses on mass semantic alignment for behaviors \cite{ji2023ai}, but there are so many other ways to conduct alignment.

The more important use is to clearly specify the technical needs of alignment technologies and workflows. As Tan Zhi-Xuan et al.\ submit, ``Reward-based alignment---and preference matching more generally---is only appropriate for AI systems with sufficiently local uses and scopes. In other words, it is adequate for only the narrow or minimalist versions of the value alignment problem, where the values and norms at stake can be summarized as a reward function specific to the system's scope'' \cite{zhi2024beyond}.

For example, different competences of alignment may lead to different kinds of data requirements and optimization algorithms. Knowledge is often best learned by generating a lot of questions and answers about content rather than simply feeding in the plain content \cite{SudalairajBPXCS2024}. However, for skills, many examples and repetition is more effective. Behaviors are better learned as preferences supplemented with scenarios in which those behaviors occur \cite{padhi2024value}. It is important to differentiate the transience of alignment because while semantic alignment implies (full) fine-tuning a model in advance and leaving it that way, episodic alignment needs to be done at inference time based on the context. Episodic alignment may be possible with adapters learned using parameter efficient fine-tuning methods that can be applied on the fly. The audience of alignment determines whether the procedure is bidirectional or unidirectional, as well as the source and format of the KSBs.

The discussion in this paper does not detail choosing the right scope of alignment, but such a process is clearly necessary. It may be a human-centered activity that aims for a reflective equilibrium in which the scope has a competence, transience, and audience with  fairly small internal conflict of values \cite{Moller2016,zhi2024beyond}. We will find an appropriate scope through such an approach; we will also implicitly deal with the problem of conflicting values because the reflective equilibrium will have resolved conflicts by narrowing the scope until there are few remaining.

\section{Related Work}
Alignment beyond mass semantic alignment for behaviors is not a white space, and there are plenty of theory and methods that exemplify other scopes. We recount a non-exhaustive list for illustration purposes only. Along the audience dimension, personalized alignment and mutual theory of mind take us toward dyadic and small group alignment \cite{kirk2024benefits,wang2024theory}. Alignment from unstructured text takes us toward public alignment \cite{padhi2024value,achintalwar2024alignment}. Along the transience dimension, we have already discussed how low-rank adapters can contribute to episodic behaviors \cite{hu2021lora}. For knowledge, there are emerging LLM architectures with episodic memory \cite{das2024larimar}. Along the competency dimension, we are seeing the emergence of some alignment methodologies focused on knowledge and skills \cite{li2024banishing,SudalairajBPXCS2024}.

\section{Conclusion}

In this paper, we submit a broader conception of LLM alignment that is organized around three scopes: competence, transience, and audience. We do so not only to point out that the literature is mostly focused on mass semantic alignment for behaviors, but also to delineate some of the technical methodologies that are required to pursue other scopes. We believe these methodologies to be a prerequisite for further work in pluralistic alignment; steerable pluralism can mediate between different scopes \cite{sorensen2024roadmap}, but they need to have been defined and implemented first. Scopes of the right level may not even need mediation if they contain minimal internal conflict or are at a level in which value conflicts have already been deliberated upon.

\bibliographystyle{aaai}
\bibliography{scopes}

\end{document}